\documentclass[aps,showpacs,dvips,showkeys,preprint,preprintnumbers]{revtex4}
\usepackage{graphicx}
\usepackage{color}
\usepackage{amsmath}
\usepackage{amssymb}
\def \be  {\begin{equation}}
\def \ee  {\end{equation}}
\def \ee  {\end{equation}}
\def \bea {\begin{eqnarray}}
\def \eea {\end{eqnarray}}

\begin{document}

\preprint{ECTP-2018-03}    
\preprint{WLCAPP-2018-03}
\hspace{0.05cm}

\title{Particle Yields and Ratios within Equilibrium and Non-Equilibrium Statistics}

\author{Abdel Nasser  Tawfik}
\email{atawfik@nu.edu.eg}
\affiliation{Nile University, Juhayna Square of 26th-July-Corridor, 12588 Giza, Egypt}
\affiliation{World Laboratory for Cosmology And Particle Physics (WLCAPP), 11571 Cairo, Egypt}

\author{Hayam Yassin}
\affiliation{Physics Department, Faculty of Women for Arts, Science and Education, Ain Shams University, 11577 Cairo, Egypt}
\affiliation{World Laboratory for Cosmology And Particle Physics (WLCAPP), 11571 Cairo, Egypt}

\author{Eman R. Abo Elyazeed}
\affiliation{Physics Department, Faculty of Women for Arts, Science and Education, Ain Shams University, 11577 Cairo, Egypt}
\affiliation{World Laboratory for Cosmology And Particle Physics (WLCAPP), 11571 Cairo, Egypt}

\date{\today}

\begin{abstract}
In characterizing the yields and ratios various of well identified particles in the ALICE experiment, we utilize extensive {\it additive} thermal approaches, to which various missing states of the hadron resonances are taken into consideration, as well. Despite some non-equilibrium conditions that are slightly driving this statistical approach away from equilibrium, the approaches are and remain additive and extensive. Besides van der Waals repulsive interactions (assuming that the gas constituents are no longer point-like, i.e. finite-volume corrections taken into consideration), finite pion chemical potentials as well as perturbations to the light and strange quark occupation factors are  taken into account. When confronting our calculations to the ALICE measurements, we conclude that the proposed conditions for various aspects deriving the system out of equilibrium notably improve the reproduction of the experimental results, i.e. improving the statistical fits, especially the finite pion chemical potential. This points out to the great role that the non-equilibrium pion production would play, and the contributions that the hadron resonance missing states come up with, even when the principles of statistical extensivity and additivity aren't violated. These results seem to propose revising the conclusions propagated by most of the field,  that the produced particles quickly reach a state of  local equilibrium leading to a collective expansion often described by fluid dynamics. This situation seems not remaining restrictively valid, at very large collision energies. 
\end{abstract}


\maketitle

\tableofcontents
\makeatletter
\let\toc@pre\relax
\let\toc@post\relax
\makeatother

\section{Introduction}
\label{sec:Intr}

Our understanding of the particle production in the high-energy collisions has been improved, drastically \cite{Tawfik:2014eba}. Currently, we understand that the colliding particles (hadrons) are being smashed, enormously \cite{AndronicNature}. Before they finally recombine onto hadrons, they form a parton state, in which quarks and gluons are likely created and strongly interacting. Such a non-equilibrium state, e.g. unstable quark-gluon plasma (QGP), rapidly expands and lastly cools down that it evolves to an entirely different state; an equilibrium one, either chemically or later on thermally, for instance, into hadrons again. Thus, the produced particles, which are nothing but hadons, i.e. we started up with colliding hadrons and at the end we are left with hadrons, are the products out of a non-equilibrium state \cite{Tawfik:2014eba}. The QCD phase transition and the related symmetry breaking and/or restoration greatly contribute to the out-of-equilibrium status of the system of interest.

Due to the absence of an alternative well-functioning approach, overall chemical and thermal equilibria are assumed, for instance, in performing lattice QCD simulations. With this regard, one would recall that the hydrodynamic approachs rely on {\it local} thermodynamic equilibrium, while viscous fluid approximation considers deviation from the local equilibrium. All these approaches are not considered in the present study. Furthermore, when proposing conditions for the chemical freezeout which enable us the study of the bulk properties of yields and ratios of the produced particles, chemical equilibria should be adopted. The study of the transverse momentum spectra is almost exclusively based on thermal equilibrium, known as thermal freezeout. These assumptions seem to work well in characterizing the statistical nature of the bulk properties, at energies up to top RHIC energy. Within this energy regime, the transverse momentum spectra of precisely-detected particles, such as pions, kaons and protons, are well described by Boltzmann-Gibbs (BG) statistics. But when moving to the LHC energies, it seems that the picture of global equilibrium faces great challenges. The particle yields and ratios and their transverse momentum spectra aren't well characterized. We might highlight what is called anomalies in pion-to-proton ratios at LHC, for a recent status we refer to \cite{ppion2018}, where the inclusion of hadronic cascade after the hypersurface of the chemical freezeout inducing non-equilibrium corrections in a very simple picture contributes to the explanation of the pion-to-proton anomalies \cite{Becattini2012,Karpenko2013}. This picture was challenged, see for instance \cite{Akkelin2002,BlaschkeRatios}, that pions not protons which are anomalously created. Also, the present script aims at an unbiased understanding of the particle production at the LHC energies. Otherwise, one enforces the system towards a very biased description.

Aiming at empowering the {\it equilibrium} statistical approaches to cover the LHC energies, as well, various proposals have been made, so far \cite{andronic2012,pAnomal1,pAnomal2,pAnomal3}.  To this end, anomalies in production and annihilation of certain particles (such as proron anomaly) and/or non-extensive statistical approaches (such as Tsallis) have been proposed to be applied to LHC energies. Despite the great success of our generic (non)extensive statistical approach \cite{Tawfik:2018ahq,Tawfik:2017bsy,Tawfik:2016pwz}, we wanted to implement, in the present script, a new idea that the pions \cite{Akkelin2002,BlaschkeRatios}, the low-lying Nambu-Goldstone bosons, are the produced particles which significantly affect both bulk and flow properties of the other particles produced in collsions at LHC nergies. We recall this assumption, which originally dates to $1990$, that the production of pion particles takes place out of a non-equilibrium process \cite{ruuskanen90}. Acually, we believe that such an idea would be rooted in the pioneering works of Bogolubov on the boson superfluidity based on degeneracy of a non-perfect Bose-Einstein gas \cite{Bogolubov1947} and determining the second quantization of the energy spectrum \cite{Landau1949}. 

The model shall be outlined in the next section followed by a short review of the Hadron Resonance Gas (HRG) model in {\it equilibrium}, in which quark occupation factors are allowed to take values differ from the ones characterizing an equilibrium status. The inclusion of the repulsive van der Waals interactions, which are apparently slightly drive the system towards non-equilibrium, at least by assuming that the constituents aren't point-like, shall be discussed, as well. Then, we elaborate the modifications carried out to implement non-perfect boson (pion) gas based on Bogolubov's superfluidity, where finite pion-chemical potential could be imposed \cite{Bogolubov1947}. Assuming finite pion chemical potential implies the existence of Bogoliubov dispersion relation for the low-lying elementary excitations of the pion fluid; boson fluid. On the other hand, when the state of statistical equilibrium is degenerate, the degeneracy of the equilibrium states is removed adding a noninvariant term to the Hamiltonian, such as pion chemical potential.

Finally we discuss on the results obtained and draw the final conclusions, which are highlighted with the ones stemming from generic (non)extensive statistical approaches.

\section{Model}
\label{sec:mdl}

\subsection{Hadron Resonance Gas with van der Waals Interactions}
\label{sec:hrgEqlbm}

If the hadron resonances are treated as an ideal (non-interacting) gas, the equilibrium thermodynamic pressure of the hadronic phase of the Quantum Chromodynamics (QCD) can be very well determined and successfully confronted to the lattice QCD calculations \cite{Karsch:2003vd,Karsch:2003zq,Redlich:2004gp,Tawfik:2004sw,Tawfik:2004vv,Tawfik:2004vvB,Tawfik:2004vvC,Vunog}. For details, interested readers are kindly advised to consult the most recent review article \cite{Tawfik:2014eba}. It is apparent that the HRG model is based on an additive principle, i.e. Boltzmann-Gibbs (BG) extensivity,  
\bea 
& &\ln Z(T, \mu_h,V) = V \sum_h\pm \frac{g_h}{2\pi^2}\int_0^{\infty} k^2 dk \times \nonumber \\
& & \hspace*{10mm} \ln\left\{1 \pm \left.\gamma_l^{nq_l}\right|_h \cdot \left.\gamma_s^{nq_s}\right|_h \cdot \exp\left[\frac{\mu_h -\varepsilon_h}{T}\right]\right\}, \;\;\; \label{eq:lnz1}
\eea
where $\varepsilon_h=(k^2+ m_h^2)^{1/2}$ is the $h-$th particle dispersion relation, $g_h$ is spin-isospin degeneracy factor and $\pm$ stands for fermions and bosons, respectively. $\gamma_l$ and $\gamma_s$ stand for the occupation factor of light and strange quarks, respectively \cite{Rafelski2018,Becattini2004}. The quark composition of a given hadron is apparently encoded in $nq_l$ and $nq_s$ for light and strange quarks, respectively. This should be also reflected in the corresponding chemical potentials of the various quantum charges, such as baryon, strangeness, electrical charge, etc. At equilibrium, both $\gamma_l$ and $\gamma_s$ are assigned to unity allowing to omit both quantities $\left.\gamma_l^{nq_l}\right|_h$ and $\left.\gamma_s^{nq_s}\right|_h$ for $h$-th hadron from front of the exponential function. At non-equilibrium, the values of both factors (or any of them) read $\left.\gamma_l^{nq_l}\right|_h\neq 1$ and $\left.\gamma_s^{nq_s}\right|_h\neq 1$ \cite{Tawfik:2005gk}.

Contributions from all known resonance states with mass $\leq 2~$GeV \cite{hgdrn,hgdrnB} taken from recent compilation of the particle data group (PDG) \cite{pdg2018} can be summed up, additively. Such a mass cut-off defines the temperature-validity of the HRG model, setting a natural limitation to the hadronic phase. Furthermore, we take into consideration the proposal that the so-called missing states should be included, as well \cite{Capstick1986}. Concretly, we mean with missing states the baryons predicted in ref. \cite{Capstick1986} and not yet confirmed, experimentally \cite{pdg2018}. They are entering our calculations in the same matter as we do for the PDG hadrons and resonances \cite{Tawfik:2014eba}. In other words, this procedure shouldn't violate the additivity and extensivity statistical principles. The missing states are resonances predicted, theoretically, but not yet confirmed experimentally. Their quantum numbers and physical characteristics are theoretically well known. Basically, they are conjectured to greatly contribute to the fluctuations and the correlations estimated in recent lattice QCD simulations \cite{Karsch2014}. These were the resons why to add them \cite{Redlich2016}. Despite the convention that this wouldn't be the case in the present study, as we are strictly focusing on particle yields and ratios, we wanted to add them, as the particle ratios, for instance, aren't entirely free of correlations \cite{Alba2017}. Another reason for adding the missing states is that they come up with additional degrees of freedom and considerable decay channels to the particles which are subject of this present study.

As given earlier, the constituents of HRG are free (collisionless) particles. Some authors prefer taking into account the repulsive van der Waals interactions in order to partly compensate the strong interactions in the hadronic medium \cite{Tawfik:2013eua} and/or to significantly drift the system towards a non-equilibrium status.  Accordingly, each constituent is allowed to have an {\it eigen}volume and the resulting hadron system becomes thermodynamically non-ideal non-equilibrium.  The repulsive interactions between hadrons are considered as a phenomenological extension and exclusively based on van der Waals excluded volume \cite{exclV1,exclV2,exclV3,exclV4}. Thus, the resulting total volume might be subtracted from the entire volume of the fireball. Considerable modifications in the thermodynamics of the hadron gas such as energy, entropy and number densities are likely expected. The hard-core radius of hadron nuclei can be related to the multiplicity fluctuations and consequently the particle yields and ratios.

Ability of the HRG model with finite-volume constituents without missing states in reproducing the lattice QCD thermodynamics has been confirmed, long ago \cite{Tawfik:2013eua}. But which limits should be set to the proposed {\it eigen}volume? To this end, few remarks are now in order. At radius $r>0.2~$fm, the disagreement with these first-principle calculations, the lattice QCD, is very convincing. Such an agreeent become more and more excellent  with the increase in the particle radii. But, at higher temperatures, the resulting thermodynamic quantities turn to be {\it non}-physical. Thus, it was concluded that the excluded volume correction becomes practically irrelevant, as it comes up with a negligible effect at $r\leq 0.2~$fm. On the other hand, a remarkable deviation from the lattice QCD calculations appears, especially when relative large values are assigned to the hadron radii. Intensive theoretical works have been devoted to the estimation of the excluded volume and its effects on the particle production and fluctuations, for instance \cite{exclV5}. It is conjectured that the hard-core radius of hadron nuclei can be related to the multiplicity fluctuations and as a consequance to the particle yields and ratios\cite{exclV6}. In the present work, we simply assume that all hadrons are spheres and  all have same radius. On other hand, the assumption that the radii would depend on the hadron masses and sizes could come up with a very small improvement. Therefore, we can neglect this.

It is obvious that various types of interactions should be assumed, as well \cite{Stoecker2018,Stoecker2018B}. For a possible inclusion of the strong interactions, themselves, we advise intereted readers to recall ref. \cite{Tawfik:2004sw}, where a pioneering theoretical estimation was introduced. Nevertheless, we limit our calculations here to the van der Waals repulsive interactions, which could be estimated by replacing the system volume $V$ by an actual one $V_{act}$, 
\begin{eqnarray}
V_{act} &=& V - \sum_h\, v_h\, N_h, \label{eq:evc1}
\end{eqnarray}
where the volume and particle number of each constituent hadron, respectively, read $v_h=4\, (4 \pi r_h^3/3)$ and $N_h$. The {\it eigen}volume $v_h$ characterized the $h$-th hadron particle.  From Eq. (\ref{eq:evc1}), we get $\tilde{\mu}_h=\mu_h - v_h\, p$, where the thermodynamic pressure $p$ should be determined in a self-consistent matter $\sum_h p_h^{id}(T,\tilde{\mu}_h)$ and 
\begin{eqnarray}
n &=& \frac{\sum_h n_h^{id}(T,\tilde{\mu}_h)}{1+\sum_h v_h n_h^{id}(T,\tilde{\mu}_h)}, \\
\epsilon &=& \frac{\sum_h \epsilon_h^{id}(T,\tilde{\mu}_h)}{1+\sum_h v_h n_h^{id}(T,\tilde{\mu}_h)}, \\
s &=& \frac{\sum_h s_h^{id}(T,\tilde{\mu}_h)}{1+\sum_h v_h n_h^{id}(T,\tilde{\mu}_h)}
\end{eqnarray}
The superscript $id$ refers to point-like calculations.
  
In the section that follows, we discuss on out-of-chemical equilibrium based on finite pion chemical potential that we are proposing to be inserted into the partition function.

\subsection{Non-Equilibrium Approaches} 
 
\label{sec:NoNhrgEqlbm}

At out-of-chemical equilibrium and inspired by the proposal of Bogolubov for superfluidity as degeneracy of a non-perfect Bose-Einstein gas and a general form of the energy spectra; a kind of ingenious application of second quantization, a thermal distribution was proposed in ref. \cite{ruuskanen90}. The pion condensation can be seen as a form of Bogolubov superfluidity. It was assumed that the matter of interest, in our case the QCD matter, is formed as a cylindrical tube with a radius $r$ and rapidly longitudinally expands without transverse flow $\nu_z=z/t$.

This approach was successfully implemented in characterizing the $p_T$-spectra of th {\it negatively} charged bosons from the $200~$A~GeV O+Au and S+S collisions in the NA35 experiment \cite{ruuskanen90}. To this end, $\varepsilon$ in Eq. (\ref{eq:lnz1}) should be replaced by the azimuthal angle $\phi$ and the covariant form $p_{\mu}\, u^{\mu}$ (four-momentum and -velocity) and then integrating every thing over the freeze-out time $\tau_{fo}=\tau$; the pion transverse mass reads $m_T=(p_T^2+m^2)^{1/2}$, the volume element $d^3 p$ is given in the transverse momentum $p_T$, the rapidity $y$ and the azimuthal angle $\phi$; $d^3 p=p_T m_T \cosh(y) d p_T\, d y\, d \phi$. Thus, the energy can be expressed as $\varepsilon=m_T\, \cosh(y)$.

The scalar field $\phi(x)$ is conjectured to have a unitary transformation by the phase factor $\exp(-i \alpha)$. Therefore, the Bose-Einstein condensation of the lowest-lying Nambu-Goldstone bosons can be studied in U($1$) global symmetry \cite{kapusta}. The single-particle partition function is then given as
\bea
\ln z(T,\mu_{\pi}) &=& \frac{V}{T} \left(\mu_{\pi}^2-m^2\right) \xi^2 \nonumber \\
&-& V \int \frac{d^3 p}{(2\, \pi)^3} \left[\frac{\varepsilon}{T} + \ln\left(1 \pm \gamma_l^{nq_l} \gamma_s^{nq_s} e^{-\frac{\varepsilon-\mu_{\pi}}{T}}\right) \right].
\eea
The parameter $\xi$, which can be treated as a variational parameter relating to the charge of condensed particle, carries the infrared characters of the scalar field. At $|\mu_{\pi}|<m$, the equilibrium partition function, Eq. (\ref{eq:lnz1}), can be recovered.

\section{Results}
\label{sec:Rslts}

\begin{figure}[htb]
\includegraphics[width=8cm,angle=-0]{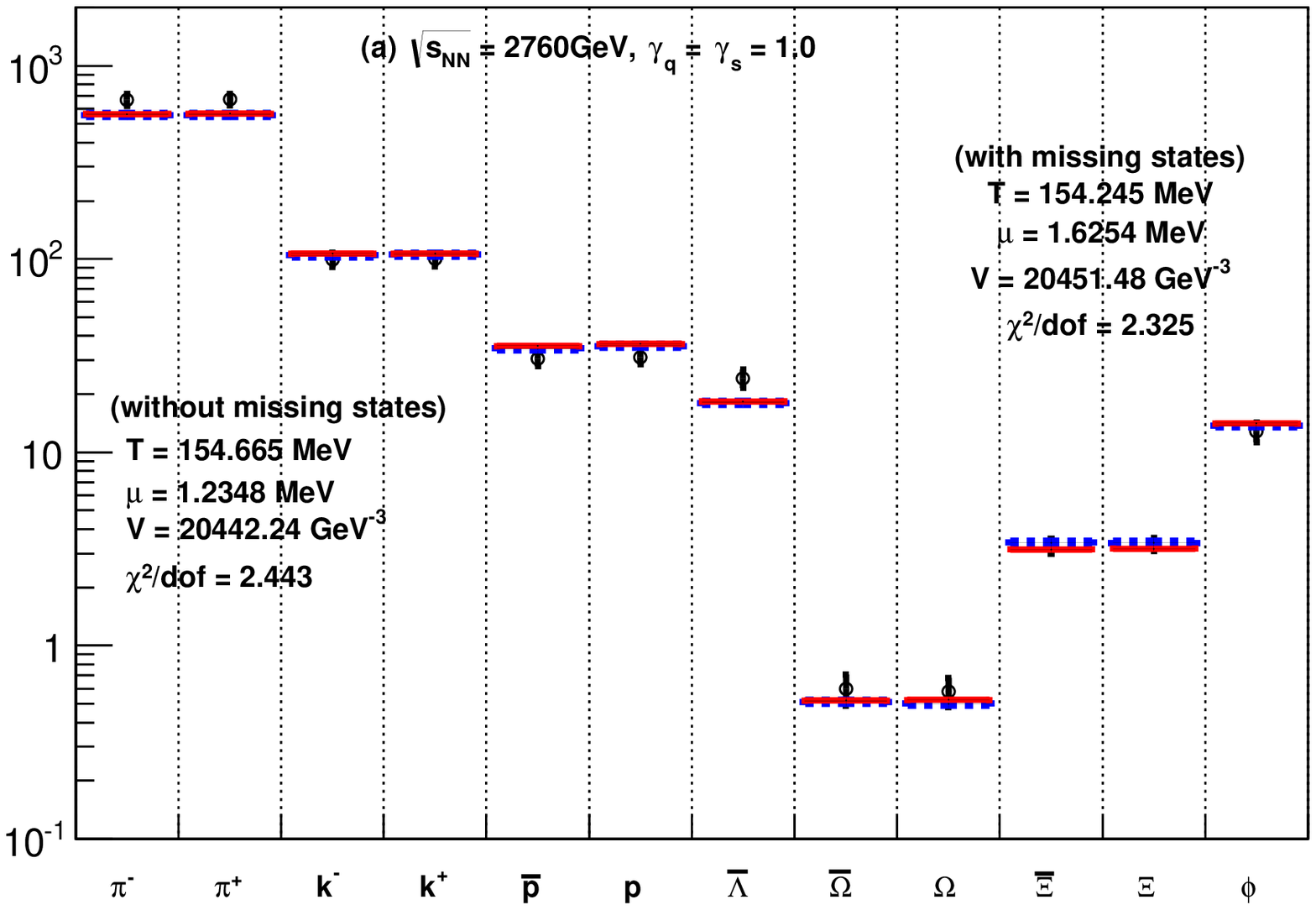}
\includegraphics[width=8cm,angle=-0]{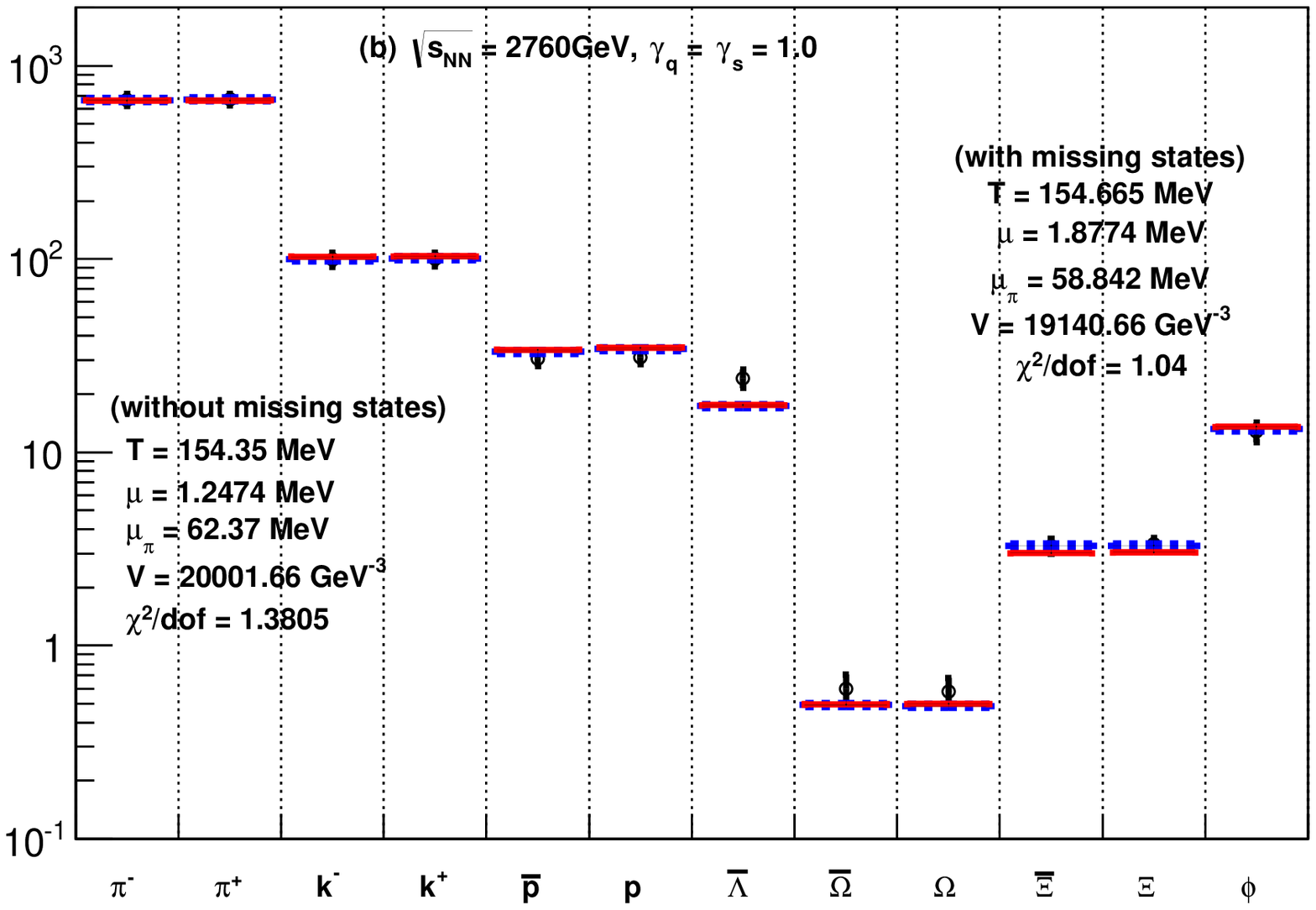}
\caption{Variuous particle yields calculated from the statistical thermal model with (dashed lines) and without missing states (solid lines) at equilibrium $\gamma_l=\gamma_s=1$ are fitted to ALICE measurements (symbols). The top and bottom panels compare between results at vanishing $\mu_{\pi}$ (top panel) and finite $\mu_{\pi}=0$ (bottom panel). \label{fig:yls1} }
\end{figure}

\begin{figure}[htb]
\includegraphics[width=8cm,angle=-0]{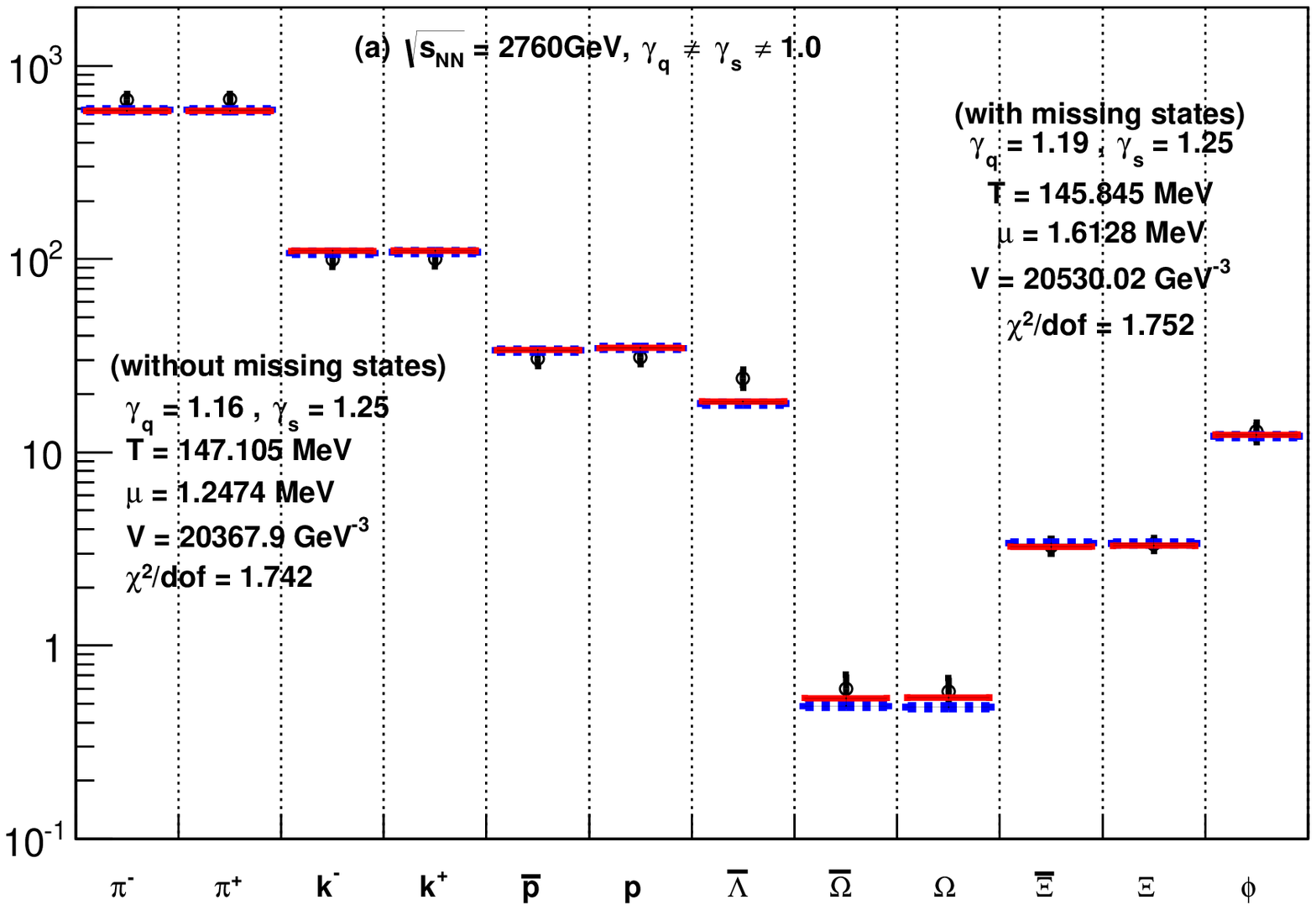}
\includegraphics[width=8cm,angle=-0]{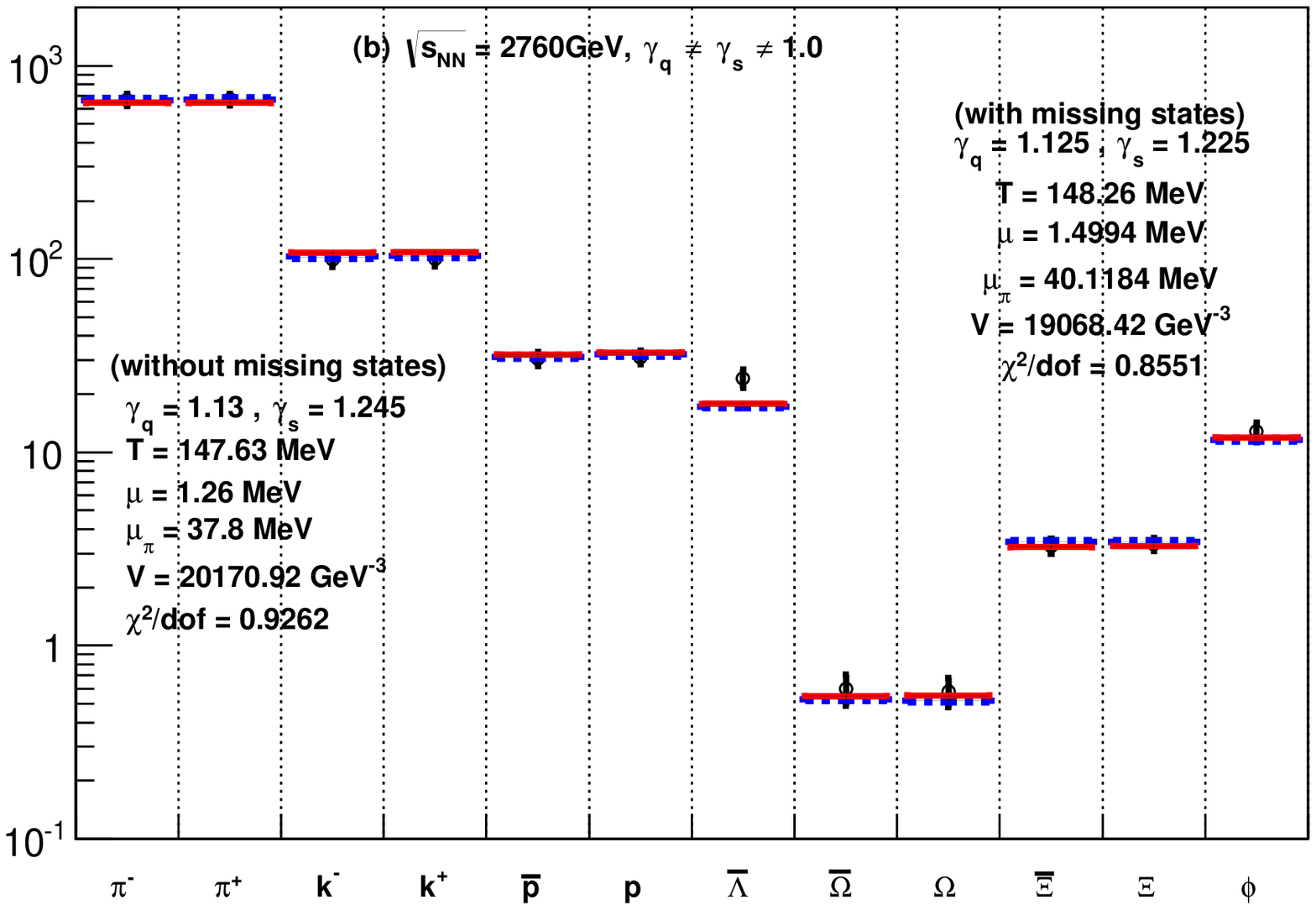}
\caption{The same as in Fig. \ref{fig:yls1} but $\gamma_l$ and $\gamma_s$ are taken as free fitting parameters.\label{fig:yls2} }
\end{figure}

The results on the particle yields $\pi^+$, $\pi^-$, $\mathtt{K}^-$, $\mathtt{K}^+$, $\bar{\mathtt{p}}$, $\mathtt{p}$, $\bar{\Lambda}$, $\bar{\Omega}$, $\Omega$, $\bar{\Xi}$, and $\phi$ calculated from the thermal model, Sec. \ref{sec:hrgEqlbm}, with (dashed lines) and without missing states (solid lines) at equilibrium $\gamma_l=\gamma_s=1$ fitted to the ALICE measurements (symbols) are depicted in Fig. \ref{fig:yls1}. The experimental results are given as symbols with errorbars. The top and bottom panels present a comparison between our calculations at vanishing $\mu_{\pi}$ (left) and finite $\mu_{\pi}=0$ (right).  The resulting parameters are listed inside the graphs. From the given $\chi^2/\mathtt{dof}$, we can draw a conclusion that adding missing state to the PDG compilation considerably improves the statistical fits. 

The results at $\gamma_l\neq 1$ and $\gamma_s\neq 1$ are depicted in Fig. \ref{fig:yls2}. When analyzing the resulting fitting parameters, we can draw another conclusion that when $\gamma_l$ and $\gamma_s$ are allowed to take values differ from unity, the ability of this {\it partly} non-equilibrium to reproduce the experimental data increases and this considerably improves the statistical fits. When comparing both figures \ref{fig:yls1} and \ref{fig:yls2} and the resulting fitting parameters we conclude that the pion chemical potential excellently describe the particle yields at the LHC energy.

\begin{figure}[htb]
\includegraphics[width=8cm,angle=-0]{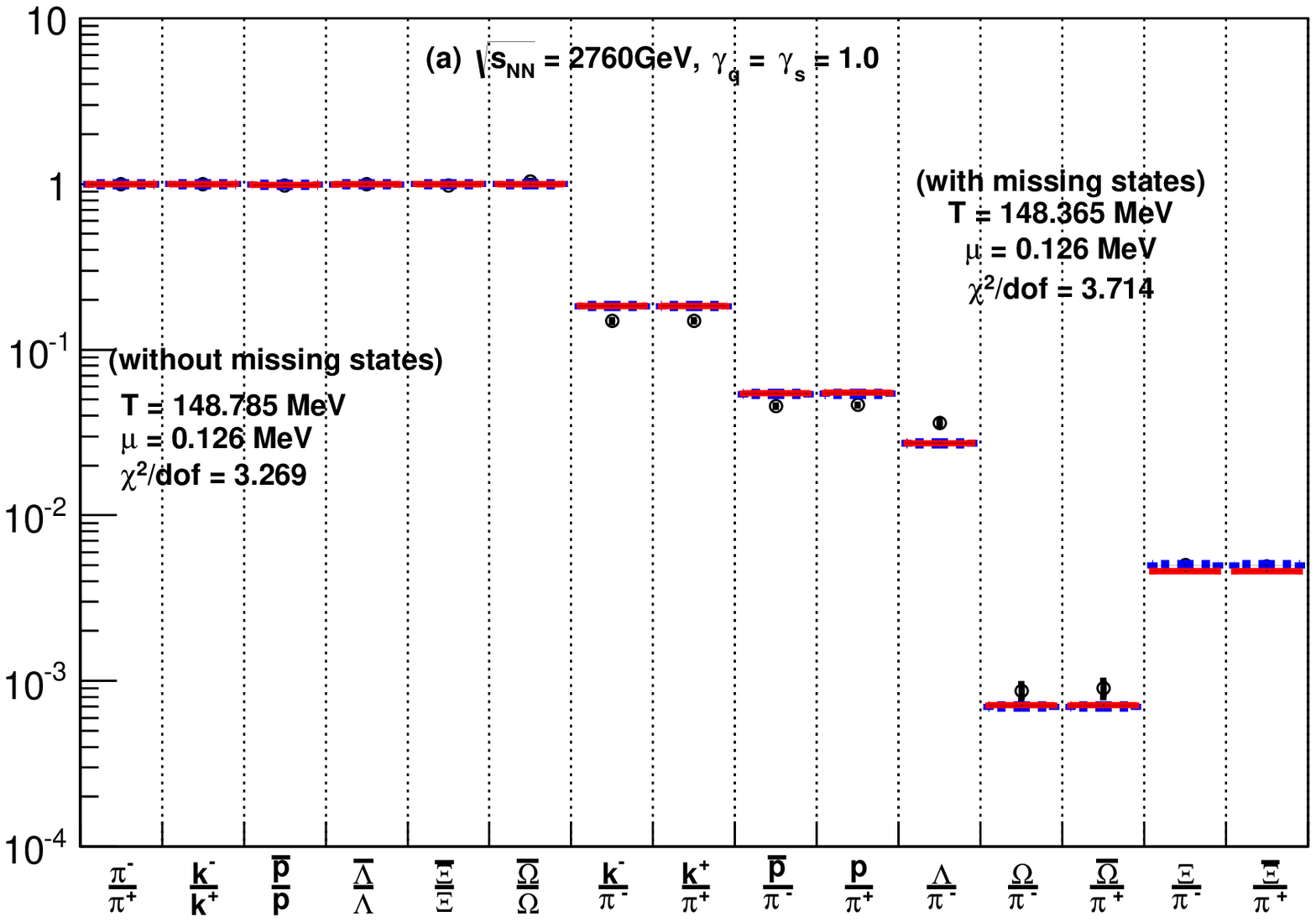}
\includegraphics[width=8cm,angle=-0]{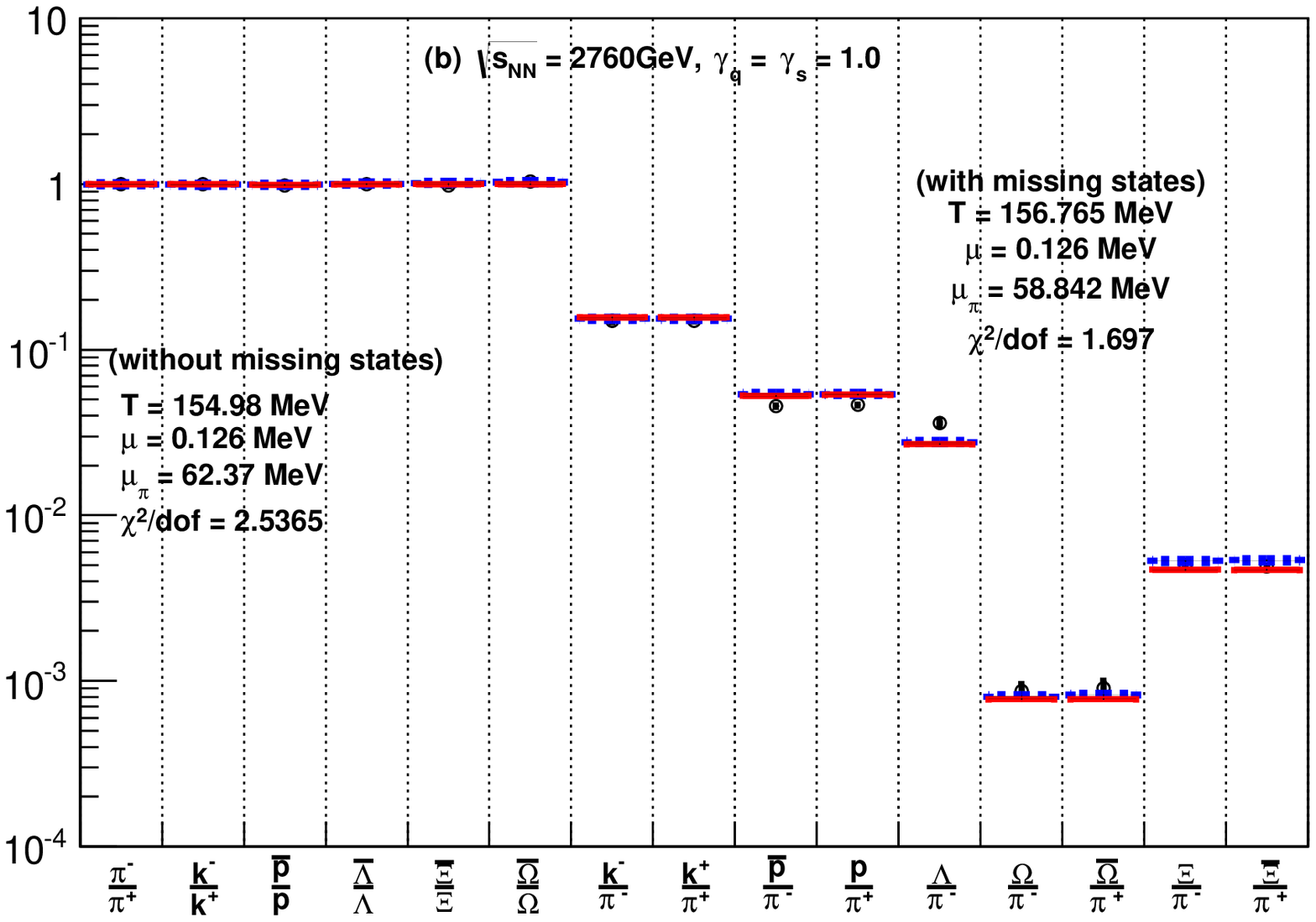}
\caption{The same as in Fig. \ref{fig:yls1} but here for the particle ratios at $\gamma_l=\gamma_s=1$.\label{fig:rts1} }
\end{figure}

\begin{figure}[htb]
\includegraphics[width=8cm,angle=-0]{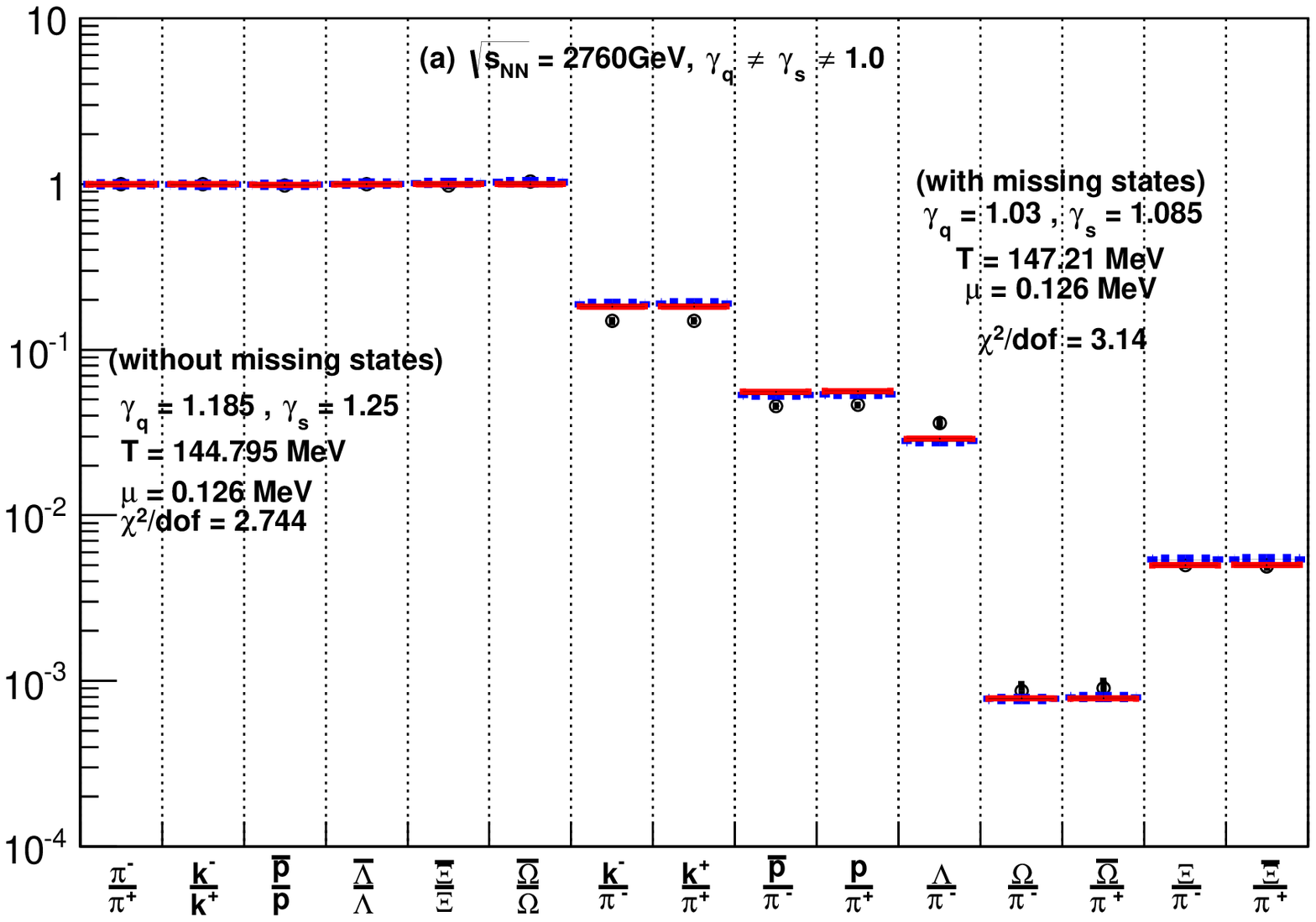}
\includegraphics[width=8cm,angle=-0]{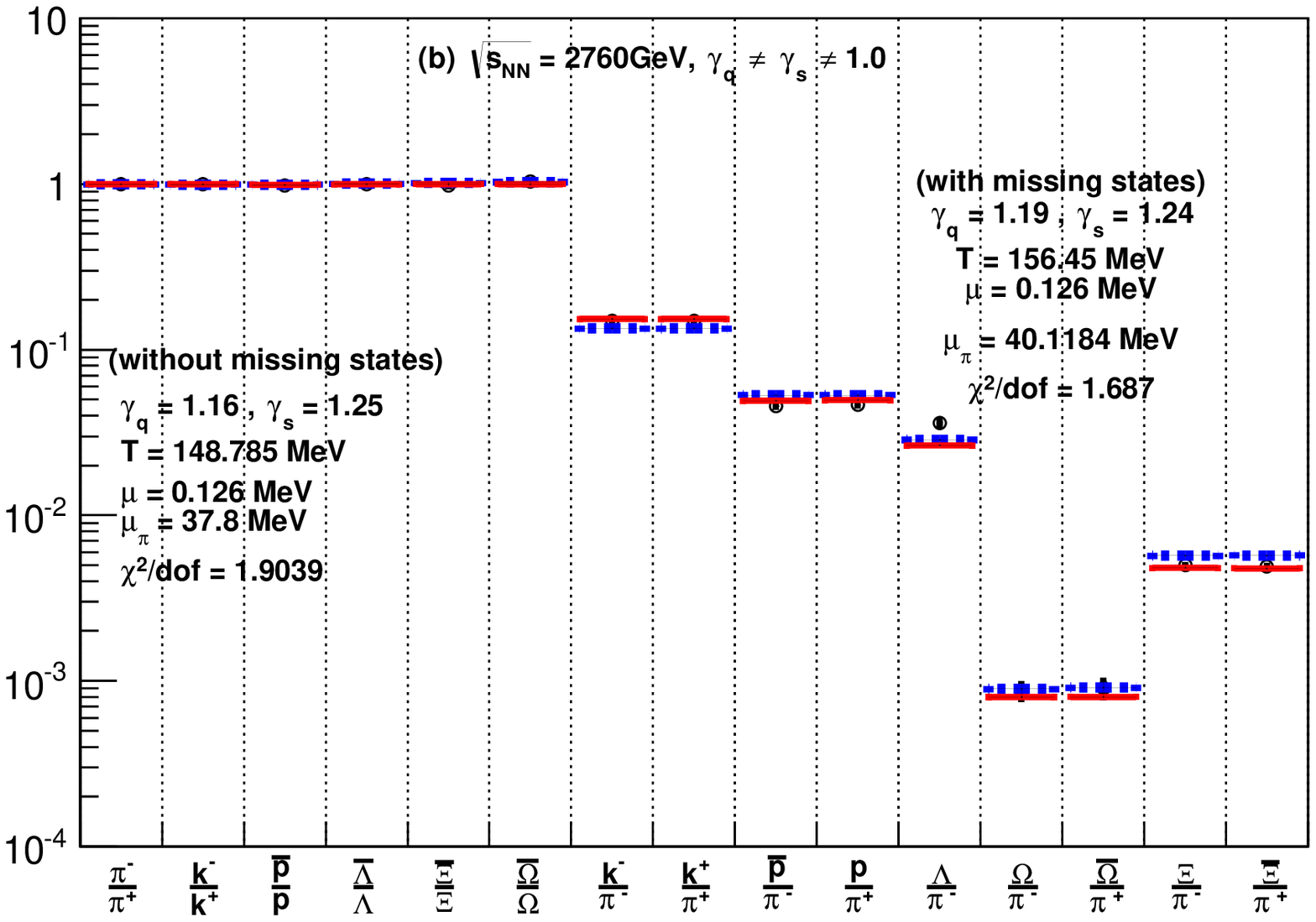}
\caption{The same as in Fig. \ref{fig:rts1} but here $\gamma_l$ and $\gamma_s$  are allowed to take values differ from unity. \label{fig:rts2} }
\end{figure} 

Similar to figures \ref{fig:yls1} and \ref{fig:yls2}, the results on the particle ratios $\pi^-/\pi^+$, $\mathtt{K}^-/\mathtt{K}^+$, $\bar{\mathtt{p}}/\mathtt{p}$, $\bar{\Lambda}/\Lambda$, $\bar{\Omega}/\Omega$, $\bar{\Xi}/\Xi$, $\mathtt{K}^-/\pi^-$,  $\mathtt{K}^+/\pi^+$, $\bar{\mathtt{p}}/\pi^-$, $\mathtt{p}/\pi^+$, $\Lambda/\pi^-$, $\Omega/\pi^-$, $\bar{\Omega}/\pi^+$, $\Xi/\pi^-$, and $\bar{\Xi}/\pi^+$ are depicted in Figs. \ref{fig:rts1} and \ref{fig:rts2}, respectively. The resulting parameters are listed out inside the graphs. This time, the fireball volume isn't included in. The particle ratios are likely cancel the dependence on $V$, at least to very large extent. The volume fluctuations, on the other hand, might not be entirely removed. This might be the statistical price to be paid, as long as, not other alternative exists so-far. Another conclusion can be drawn here, as well. $\mu_{\pi}\neq 0$ helps in improving the reproduction of the different particle ratios by means of the statistical thermal approaches. 

With this regard, we might recall a recent study on production of $\pi$,  $\mathtt{K}$,  $\mathtt{p}$, and  $\Lambda$  and their ratios in Pb+Pb collisions at $2.76~$TeV in blast-wave model with thermal equilibrium mechanism \cite{PRC2014}.  While the antiparticles-to-particles  and the kaons-to-pions ratios were well reproduced, the  $\mathtt{p}/\pi$ was overestimated by a factor of $1.5$. Based on this study \cite{PRC2014}. it was found that  $\mathtt{p}/\pi$ and $\mathtt{K}/\pi$ are dominated by the radial flow. In the present study, we have, among others, avoided constraining the fitting parameters as done in ref. \cite{PRC2014}. To solve the $\mathtt{p}/\pi$-overestimation, other proposals have been discussed in Sec. \ref{sec:Intr}.

When comparing our results with the THERMUS predictions, we conclude that our fitting parameters are smaller. While our strangeness chemical potential $\mu_S$ is determined at $T$ and $\mu_B$ to assure strangeness neutrality, the resulting fitting parametersl, at $\gamma\neq 1$, $\gamma_s\neq 1$ with missing hadron states, read $\mu_B=0.126\pm 0.01~$MeV (in THERMUS $\mu_B$ is fixed to $1~$MeV), $\gamma=1.19\pm 0.05$, $\gamma_s=1.24\pm 0.07$, $\mu_{\pi}=40.12\pm0.65~$MeV, and $T=156.45\pm 1.75~$MeV. 

\section{Conclusions}
\label{sec:Cncl}

In reproducing the measured particle yields and ratios at the LHC energies, we decided in favor for a new  alternative approach, namely the Bogolubov superfluidity of Bose-Einstein gas, i.e. the formation of pion condensation as $\mu_{\pi}\rightarrow m_{\pi}$ could be regarded as analogy of the Bogolubov superfluidity. We wanted to show whether the non-equilibrium pion production (associated with $\mu_{\pi}\neq0$) affects other bulk properties at the LHC energies, such as the particle numbers. In addition to this, we have taken into consideration various missing states of the hadron resonances. In additional to the baryon, the strangeness  and the electric charge potentials, we have imposed $\mu_{\pi}\neq0$, as well. The extensive HRG model is extended to finite-volume constituents and the  light and strange quark occupation factors are taken at equilibrium as well as at non-equilibrium. 

Our measure for the best statistical fitting was the smallest $\chi^2$/dof. In fitting particle yields, we should/could estimate the fireball volume. We noticed that best fits are accompanied by smallest volume and lowest $\mu_{\pi}$, as well. Accordingly, we were able to draw the conclusions that the various particle yields and ratios produced in the most central ($0-5\%$) Pb+Pb collisions at $2.76~$TeV are well reproduced at $\mu_{\pi}\neq 0$ while both $\gamma_l$ and $\gamma_s$ take values greater than unity. 

The present study shows that the particle yields and ratios at the LHC energy would be best reproduced in the extensive thermal models, as well done lower energies. This is conditioned to pion production anomaly, i.e. finite $\mu_pi$. The reasons why this appears at LHC energy should be subject of future works. The attempt introduced in ref. \cite{Akkelin2002,BlaschkeRatios}, should be intensified. 

Various mechanisms for non-equilibrium particle prodcation should be integrated in the thermal approachs. In the present work, we have avoid to implement non-extensive statistical approaches even in this relativistic energy regime, especially that the Tsallis-type approaches are not fully correctly applicable at lower energies. Such a conclusion seems to confirm an old study of AT, where he utilized a  generic (non)extensive statistics \cite{Tawfik:2017bsy}. The latter is assumed to determine the degree of nonextensivity based on two equivalent classes $(c,d)$, where BG is related to $(c,d)\equiv(1,1)$ and Tsallis to $(c,d)\equiv(q,0)$. The physical meaning of these two classes and their relation to (non)equilibrium mechanism are discussed in ref. \cite{Tawfik:2018ahq}. If their values differ from unity, $(c,d)$ can be related to Lambert-$W$ exponentials characterizing the entropic equivalence classes. Thus, $(c,d)$ are scaling exponents. For example, $c=1$ and $d>0$ give a fractional power law and the entropy of the system of interest is characterized by delayed relaxation. If $(c,d)$ fulfill Shannon-Khinchin axioms on continuity, maximality, expandability, and generalized additivity, they are charactering extensivity. But when violating the fourth axiom, even individually, $(c,d)$ manifest nonextensive entropy. Based on the generic (non)extensive statistics various particle yields at top and lowest RHIC-energies were well reproduced by BG rather than any other nonextensive approach. The resulting scaling exponents are close to stretched exponentials, i.e. $(c,d)\equiv(1,d>0)$ and asymptotically stable classes of entropy. This means that 
\bea
S_{\eta}(p) &=& \sum_i \Gamma\left(\frac{\eta+1}{\eta},-\ln p_i\right) - p_i \Gamma\left(\frac{\eta+1}{\eta}\right),
\eea
where in this particular case the equivalent class $\eta=1/d$ is characterized as stretching exponent distribution, i.e. $\eta>0$ \cite{AP1999}. 

At $(d\equiv 1/\eta)>0$, the branch of Lambert-$W$ functions, which are the real solutions of $x=W_k(x) \exp(W_k(x))$ are the ones within $k=0$, $W_0(x)\sim
x-x^2+\cdots$. Within the given $\eta$-region, three cases can be identified: \begin{itemize}
\item At $\eta<1$, $S_{\eta}$ is superadditive.
\item At $\eta>1$, $S_{\eta}$ subadditive. 
\item At $\eta=1$, $S_{\eta}$ is characterized by positivity, equiprobability, concavity and irreversibility. This means that the first three Shannon-Khinchin axioms (continuity, maximality, and expandability) besides extensivity are verified. This is nothing but the {\it logarithmic} BG nonextensive entropy.
\end{itemize} 
As example, at  $(c,d)\equiv(1,2)$, where $\eta=1/2$, we get 
\bea
S_{1,2}(p) &=& 2 \left(1-\sum_i p_i \ln p_i\right) + \frac{1}{2} \sum_i p_i \left(\ln p_i\right)^2, 
\eea
i.e. a superposition of two entropy terms. It is apparent that $S_{1,2}(p)$ is superadditive and its asymptotic behavior is dominated by the second term.

The stretched exponent distributions are characterized by $c\rightarrow 1$, where BG extensive entropy is recovered at $d=1$ and obviously the Tsallis nonextensive entropy becomes dominant, at $d=0$. The Lamber exponential reads
\bea
\lim_{c\rightarrow 1} \varepsilon_{c,d,r}(x) &=& \exp\left\{-dr\left[\left(1-x/r\right)^{1/d}-1\right]\right\},
\eea
where $r=(1-c+cd)^{-1}$ determining the size of the distribution function, especially at small probabilities of microstates ($x$) but not effecting the asymptotic properties.

\section*{Acknowledgment}

AT is very grateful to David Blaschke and Ernst-Michael Ilgenfritz for the inspiring discussions on the non-equilibrium pion production at LHC energies!

\end{document}